\documentclass[doublecol]{epl2} 

\usepackage[T1]{fontenc}
\usepackage[latin9]{inputenc}
\usepackage{amsmath}
\usepackage{graphicx}

\title{Classification of Critical Phenomena in Hierarchical Small-World Networks }

\author{S.~Boettcher and C.~T.~Brunson}

\institute{Dept.~of Physics, Emory University, Atlanta, GA 30322; USA}

\abstract{
A classification of critical behavior is provided in systems for which
the renormalization group equations are control-parameter dependent.
It describes phase transitions in networks with a recursive, hierarchical
structure but appears to apply also to a wider class of systems, such
as conformal field theories. Although these transitions generally
do not exhibit universality, three distinct regimes of characteristic
critical behavior can be discerned that combine an unusual mixture
of finite- and infinite-order transitions. In the spirit of Landau's
description of a phase transition, the problem can be reduced to the
local analysis of a cubic recursion equation, here, for the renormalization
group flow of some generalized coupling. Among other insights, this
theory explains the often noted prevalence of so-called inverted Berezinskii-Kosterlitz-Thouless
transitions in complex networks. As a demonstration, a one-parameter
family of Ising models on hierarchical networks is considered. 
}


\begin{document}

\maketitle
\section{Introduction}
One of the most significant results of network science~\cite{Barabasi03}
is the realization that critical phenomena on complex networks behave
differently than those found on a lattice geometry~\cite{Barthelemy03,Boccaletti06,Dorogovtsev08}.
Before the advent of complex networks, random geometries were routinely
modeled in terms of ordinary random graphs~\cite{erdoes1960,Bollobas,Wasserman94}.
These are well-understood and synonymous with the mean-field limit
of ordinary lattices, often with little qualitative difference in
their critical behavior~\cite{Pathria}. Therefore, it came as a surprise
that real-world networks would exhibit a dramatically distinct phenomenology,
with a profound imprint of their geometry on the dynamics. What we
now call complex networks, aside from being random, possess geometries
dominated by small-world bonds and scale-free degree distributions~\cite{Watts98,Barabasi99}. These lead to novel, and often non-universal,
scaling behaviors unknown for lattices, that have changed our appreciation,
for example, of the risk of epidemics because scale-free networks
possess a vanishing threshold for percolation~\cite{Vespignani01,Balcan09}.
In turn, the ability to conceive of \emph{synthetic} phase transitions
through the manipulation or ab-initio design of network geometry is
one of the promising targets for the emerging science of meta-materials~\cite{Metamaterials06,Regensburger12}. In particular, the iterative
structure of hierarchical networks may facilitate their realization
in engineered devices to unlock and control their unconventional behaviors.
Work on percolation~\cite{Boettcher09c,Berker09,PhysRevE.82.011113,Boettcher11d,Hasegawa13c,Singh14},
the Ising model~\cite{Bauer05,Hinczewski06,Boettcher10c,BoBr11,Baek11},
and the $q$-state Potts model~\cite{Nogawa12,PhysRevLett.108.255703,Singh14b}
have shown that critical behavior, once thought to be exotic and model-specific~\cite{Dorogovtsev08}, can be categorized with the renormalization
group~\cite{Nogawa12} for a large class of hierarchical networks
with a hyperbolic structure.

The renormalization group (RG)~\cite{Wilson71,Wilson72} is a widely
used method in statistical physics that is by now found in most textbooks~\cite{Goldenfeld,Plischke94,Pathria}.
It has allowed to categorize broad classes of equilibrium systems
into enumerable sets of universality classes, each characterized by
discrete features, such as their dimension and the symmetries adhered
to by their Hamiltonians. Such universality is made possible through
the property of \emph{scaling} that is an inherent feature near critical
points~\cite{Kadanoff66}. Scaling entails that system-specific details
on the microscopic level become irrelevant, as the behavior over many
orders in the range of the interactions become self-similar. In this
framework, analogous behavior in a surprisingly wide set of phenomena,
such as the condensation of fluids, spontaneous magnetization of materials,
or the generation of particle masses in the early universe, can be
described with a few effective theories -- a major intellectual accomplishment
of modern physics~\cite{Goldenfeld}.

\begin{figure}
\includegraphics[bb=30bp 400bp 600bp 700bp,clip,scale=0.4]{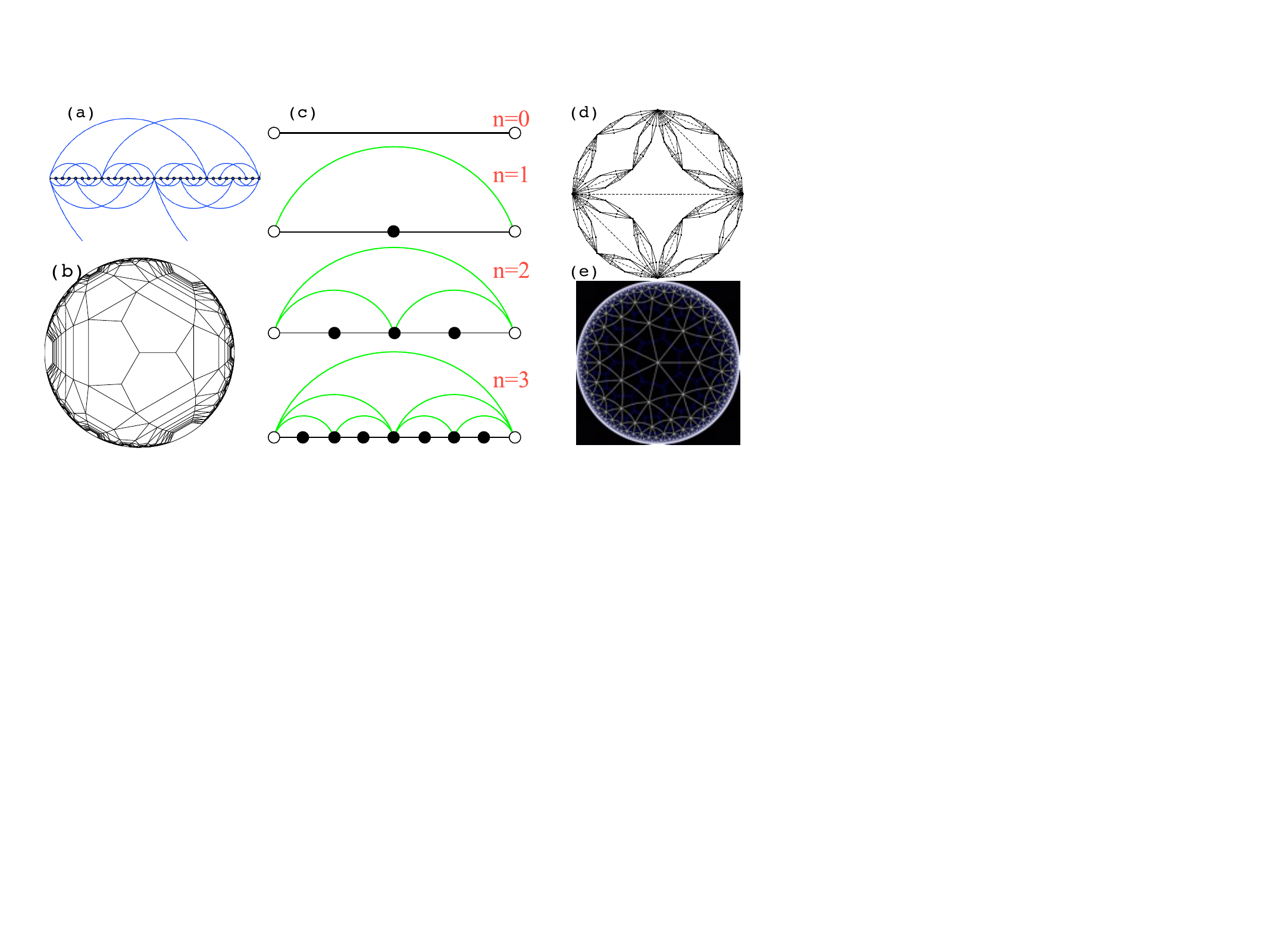}
\protect\caption{\label{fig:HierarchicalNetworks}Examples of hierarchical networks:
(a) non-planar Hanoi network~\cite{SWPRL,Boettcher09c}, (b) enhanced
binary tree~\cite{Nogawa09,PhysRevE.82.011113,Baek11}, (c) and (d)
small-world versions of the Migdal-Kadanoff RG embedded in $d=1$~\cite{Boettcher09c,Boettcher11d}
(shown for its first few recursions $n=0,\ldots,3$) and $d=2$~\cite{Hinczewski06,PhysRevLett.108.255703},
and (e) hyperbolic networks. After each recursion, small-world couplings
access ever larger pools of variables, leading to ``patchiness''~\cite{Boettcher09c}.}
\end{figure}

Unlike the Euclidean arrangement of atoms in a lattice, agents in
biological or social systems may exhibit complex networks of mutual
interactions~\cite{Watts98,Boccaletti06,Dorogovtsev08}. As the dependence
on lattice dimensionality indicates, the study of critical phenomena
is inseparable from the understanding of the geometry of the network~\cite{barthelemy_spatial_2010}.
It has been realized that many of the networks that are engineered
by some natural or human activity themselves exhibit emergent complex
properties, exemplified by the scale-free degree distribution of the
internet. While these networks, and dynamical systems on them, may
behave critical, those phenomena were soon found to be non-universal~\cite{Barabasi99,Andrade05,Auto08},
i.e., they are intimately tied to intricate details of the specific
system. In this sense, it would seem unlikely that a sweeping classification
could be devised. Here, we will categorize equilibrium phenomena observed
on a large set of networks having hierarchical structure~\cite{Andrade05,SWPRL,Hinczewski06,PhysRevE.82.011113,Baek11,Nogawa09},
as those in Fig.~\ref{fig:HierarchicalNetworks}.   Our discussion
pertains, for example, to the robustness of infinite-order transitions
in distinct network models summarized in Ref.~\cite{Dorogovtsev08},
or in field theory, where it signals the loss of conformality~\cite{Kaplan09}.  However, it is most closely related to the recent observation of discontinuous ("explosive")  transitions in ordinary percolation on hierarchical networks~\cite{Boettcher11d,Nogawa13,Singh14,Singh14b}.
Our study shows that criticality in these models is generally non-universal
but falls into three generic regimes. One of these regimes is an infinite-order
transition reminiscent of that described by Berezinskii, Kosterlitz
and Thouless (BKT)~\cite{Plischke94} but of very different origin.  Ref.~\cite{PhysRevLett.108.255703} has  provided a comprehensive scaling theory for this regimes. We find that it is flanked on one side by a transition with a weaker, algebraic
divergence, similar to a second order transition (albeit non-universal),
and on the other by a regime with an even stronger essential singularity,  with percolation as a non-generic exception. Our approach also reveals the origin of the cross-over between these regimes.

\section{Renormalization of Hierarchical Networks}
To preface our discussion, consider RG for the probability $\kappa_{n}$
of an end-to-end connection  in {}Fig.~\ref{fig:HierarchicalNetworks}(c)~\cite{Boettcher11d}.
Without the (arced) small-world bonds, recursively an infinite line
is built up with $\kappa_{n+1}=\kappa_{n}^{2}$, entailing percolation
($\kappa_{\infty}=1$) only for $\kappa_{0}=1$; any chance of missing
a bond, i.e., $\kappa_{0}<1$, looses the connection ($\kappa_{\infty}=0$).
If we now attribute a probability $p>0$ to those arcs, then $\kappa_{n+1}=p+\left(1-p\right)\kappa_{n}^{2}$
and we must distinguish two possibilities:  
\begin{enumerate}
\item If line and arc bonds
vary independently, with $p\not=\kappa_{0}$~\cite{Berker09,Hasegawa10a},
then $\kappa_{\infty}=p/\left(1-p\right)$ for $0\leq p\leq\frac{1}{2}$,
while the unstable fixed point (FP) for $p<\frac{1}{2}$ at $\kappa_{\infty}=1$
only becomes stable for $p>\frac{1}{2}$, both irrespective of $\kappa_{0}$.
A non-trivial FP $\kappa_{\infty}\left(p\right)$ that is manipulated
via an external parameter $p$ but is attained \emph{independent}
of the control-parameter, i.e., for any $\kappa_{0}<1$, is not uncommon~\cite{Snowman07},
and can lead to interesting phenomena like the cross-over between
two interchanging FP~\cite{Pelissetto02}. 
\item If, however, all bonds,
line and arc, are equivalent such that $\kappa_{n+1}=\kappa_{0}+\left(1-\kappa_{0}\right)\kappa_{n}^{2}$~\cite{Boettcher09c,Boettcher11d},
then the FP $\kappa_{\infty}=\kappa_{0}/\left(1-\kappa_{0}\right)$
\emph{explicitly} depends on the control-parameter $\kappa_{0}$.
The consequences are dramatic: $\kappa_{\infty}(\kappa_{0})$ becomes
unphysical for $\kappa_{0}>\frac{1}{2}$ where $\kappa_{\infty}=1$
is now stable, a non-trivial critical point at $\kappa_{0}=\frac{1}{2}$
ensues (that causes a \emph{discontinuous} percolation transition~\cite{Boettcher11d}),
and small-world bonds enforce sub-extensive (``patchy'') order even
for $\kappa_{0}<\frac{1}{2}$~\cite{Boettcher09c}.
\end{enumerate}

Our classification pertains to the later case,  with control-parameter dependent FP,
$\kappa_{\infty}(\kappa_{0})$. It conveniently applies to hierarchical
networks on which RG is exact and transitions can be studied in detail.
There, these regimes are characterized by the relative strength of
small-world bonds~\cite{Watts98}. Metric version of such networks,
like the Migdal-Kadanoff RG~\cite{Berker79} provide textbook examples
for RG and universality~\cite{Plischke94}. But in the advent of
complex networks, many hierarchical designs with non-metric (small-world
or scale-free) properties, like those in Fig.~\ref{fig:HierarchicalNetworks},
have been devised and studied~\cite{Andrade05,Hinczewski06,Hinczewski07,SWPRL,Nogawa09,PhysRevE.82.011113,Baek11}.

The central tenant of real-space RG consists of a procedure whereby
the partition function of the original system is mapped recursively
onto itself after tracing out a fraction $1-1/b$ of the dynamic variables,
in some form of ``blocking'' together $b$ variables. Prior couplings
$\vec{\kappa}_{n}$ between them combine non-trivially to produce
new, effective couplings $\vec{\kappa}_{n+1}$ between the remaining
variables after the $n$-th RG-step while leaving the Hamiltonian
form-invariant. This mapping constitutes the RG-flow 
\begin{equation}
\vec{\kappa}_{n+1}={\cal R}\left(\vec{\kappa}_{n}\right),\label{eq:RGflow}
\end{equation}
where ${\cal R}$ indicates a (typically non-linear) set of recursions.
In the thermodynamic limit, $n\sim\log_{b}N\to\infty$, phase transitions
are characterized purely by a local analysis for $\vec{\kappa}_{n}\sim\vec{\kappa}_{n+1}\sim\vec{\kappa}_{\infty}$
near FP of 
\begin{equation}
\vec{\kappa}_{\infty}={\cal R}\left(\vec{\kappa}_{\infty}\right),\label{eq:FPequ}
\end{equation}
independent of $\vec{\kappa}_{0}$. Here, $\vec{\kappa}_{0}$ represents
the   ``bare'' (as of yet unrenormalized) couplings of the original system. These carry the dependence
on the system's control parameter $\mu\in[0,1]$. For example, in an Ising model it may refer to the temperature via the "activity" 
$\kappa_{0}=\mu=e^{-\beta J}$ in units of $J=1$, or in a percolation model it may refer to the bond-percolation
probability, $\kappa_{0}=p=1-\mu$. Since $\vec{\kappa}_{0}$
expresses microscopic details of a potentially large family of conceivable
systems adhering to Eq.~(\ref{eq:RGflow}), the insensitivity of
$\vec{\kappa}_{\infty}$ on $\vec{\kappa}_{0}$ is an expression of
universality: only certain symmetry properties of the original systems
remain preserved by ${\cal R}$. Therefore, the linearized expansion,
$\vec{\kappa}_{n}\sim\vec{\kappa}_{\infty}+\vec{\epsilon}_{n}$ with
small $\vec{\epsilon}_{n}$ for large $n$, near the FP provides a
full accounting of the macroscopically observable properties of any
such system via the eigenvalue problem obtained from 
\begin{equation}
\vec{\epsilon}_{n+1}=\frac{\partial{\cal R}}{\partial\vec{\kappa}}\left(\vec{\kappa}_{\infty}\right)\,\vec{\epsilon}_{n}.\label{eq:linearRG}
\end{equation}
The eigenvalues $\lambda$ of the Jacobian $\frac{\partial{\cal R}}{\partial\vec{\kappa}}\left(\vec{\kappa}_{\infty}\right)$
and their eigenvectors $\vec{u}_{\lambda}$ provide the scaling exponents
and scaling fields observed in the phase transition~\cite{Pathria}.

\begin{figure}
\includegraphics[bb=20bp 70bp 340bp 330bp,clip,width=0.5\columnwidth]{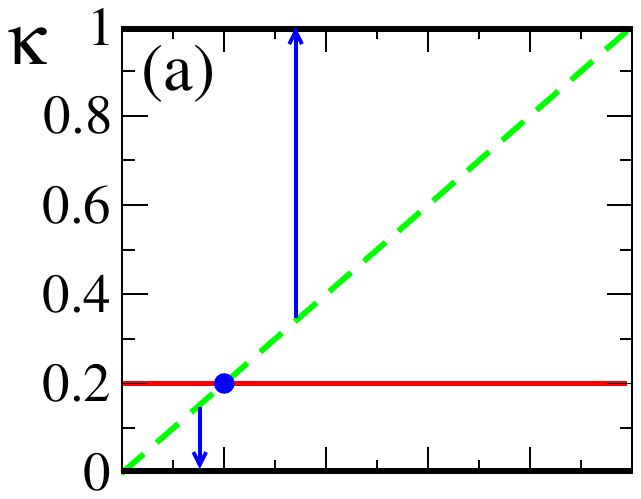}\includegraphics[bb=70bp 70bp 390bp 330bp,clip,width=0.5\columnwidth]{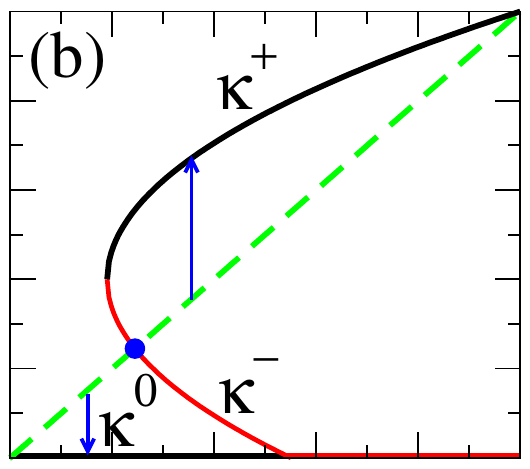}

\includegraphics[bb=20bp 20bp 340bp 320bp,clip,width=0.5\columnwidth]{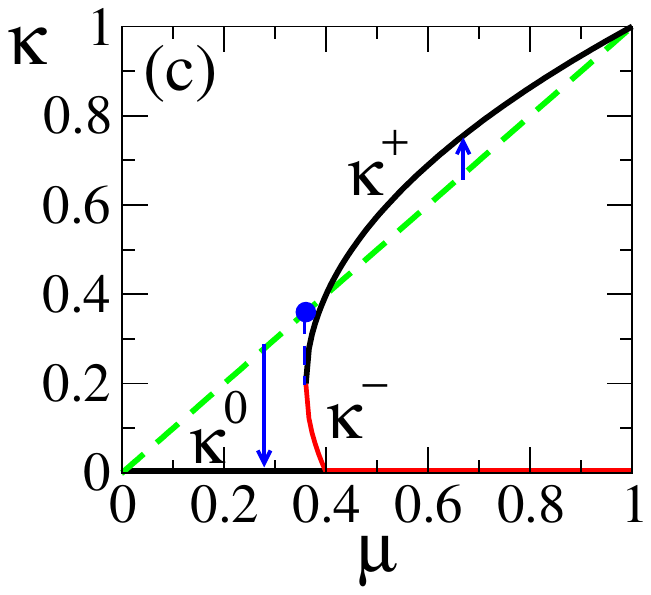}\includegraphics[bb=70bp 20bp 390bp 320bp,clip,width=0.5\columnwidth]{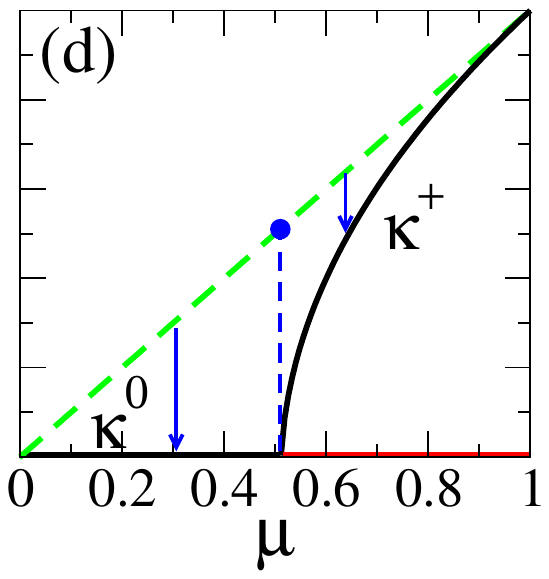}

\protect\caption{\label{fig:HN5yPlots}Generic plots of the fixed points (FP) $\kappa_{\infty}^{0,\pm}$
as a function of $\mu$ (a) for conventional FP, (b, c) for FP with
a physical branch point (BP), and (d) for FP with BP outside the physical
domain ($0\leq\mu,\kappa\leq1$). In (a), $\mu$-independence of FP
ensures universal critical behavior wherever the  bare couplings $\kappa_{0}\left(\mu\right)$
(green-dashed line) intersects the unstable FP (red line); a blue
dot and dashed line mark the critical point $\mu_{c}$, and blue arrows
indicate the RG-flow from $\kappa_{0}\left(\mu\right)$ toward the
nearest stable FP (black lines). In (b), $\kappa_{0}\left(\mu\right)$
still intersects at an unstable FP \emph{below} BP, such that the
RG-flow does \emph{not} pass BP, leading to quasi-conventional behavior
but with $\mu_{c}$-dependent critical exponents. In (c), $\kappa_{0}\left(\mu\right)$
is located above BP so that the RG-flow \emph{must} pass near BP at
$\mu_{c}$, leading to BKT-like behavior. In (d), BP drops below the
physical domain and only stable FP are accessible, resulting in an
exponential divergence at the (marginally stable) intersection $\mu_{c}$
of two FP-branches.}
\end{figure}

A much richer phenomenology arises when the RG-flow ${\cal R}$ itself
becomes dependent on the control-parameter. In that case, Eq.~(\ref{eq:RGflow})
generalizes to $\vec{\kappa}_{n+1}={\cal R}\left(\vec{\kappa}_{n};\mu\right)$
with 
\begin{equation}
\vec{\kappa}_{\infty}={\cal R}\left(\vec{\kappa}_{\infty};\mu\right)\qquad\Longrightarrow\qquad\vec{\kappa}_{\infty}=\vec{\kappa}_{\infty}\left(\mu\right),\label{eq:muRGflow}
\end{equation}
i.e., the FP $\vec{\kappa}_{\infty}$ becomes a nontrivial function
of $\kappa_{0}=\mu$. The consequences of such behavior (for a single
coupling) are depicted in Figs.~\ref{fig:HN5yPlots}. First, consider
the case of constant FP shown in Fig.~\ref{fig:HN5yPlots}(a). Drawing
constant FP as function of the control-parameter may seem redundant,
however, it allows to illustrate the connection to the bare couplings
$\kappa_{0}=\mu$ (green-dashed line). Below (above) the point where
$\kappa_{0}$ intersects the unstable FP, the RG-flow evolves toward
the stable FP at $\kappa_{\infty}=0$ ($\kappa_{\infty}=1$). Allowing
for a non-linear choice of $\kappa_{0}\left(\mu\right)$ (like, $\kappa_{0}=\mu^{y}$
for $y>0$) reflects the universality in the family of systems obeying
the same FP: no matter at which value of $\mu_{c}$ a system's  bare
coupling $\kappa_{0}\left(\mu\right)$ intersects the unstable FP,
$\kappa_{c}=\kappa_{0}\left(\mu_{c}\right)$ is always the same, which
guarantees identical (universal) critical behavior. This scenario
also applies if the RG has parameters \emph{independent} of $\mu$~\cite{Pelissetto02,Snowman07}. Like the density of long-range bonds
$p$ in our introductory example, such a parameter merely shifts the
horizontal line $\kappa_{\infty}(p)$ in Fig.~\ref{fig:HN5yPlots}(a)
up or down.

We claim that the remaining three panels in Fig.~\ref{fig:HN5yPlots}
capture \emph{all} generic features that can arise for $\mu$-dependent
FP. Variable FP $\kappa_{\infty}\left(\mu\right)$ can \emph{collide},
either linearly or in a square-root branch point (BP); any other behavior
(such as a higher-order BP) is exceptional. We can devise a simple
theory%
\footnote{Our approach is similar in spirit to Landau's model of a phase transition
found in many textbooks~\cite{Goldenfeld,Plischke94,Pathria}.%
} that reproduces these generic features. It thereby demonstrates the
generality of this classification, not only accounting for hierarchical
networks but for any physical system described by an RG-flow that
explicitly depends on its control parameter~\cite{Kaplan09,Boettcher10c,Andrade05,Hinczewski06,Hinczewski07,SWPRL,Nogawa09,PhysRevE.82.011113,Baek11}.
For example, the networks in Fig.~\ref{fig:HierarchicalNetworks}
retain memory through ever longer non-renormalizing small-world bonds
entering the flow at each level.

It is sufficient to consider the RG recursion for a single coupling
$\kappa_{n}$ with some control parameter $\mu$. We argue that FP
in a real, $\mu$-dependent RG-flow ${\cal R}$ in Eq.~(\ref{eq:muRGflow})
will exhibit BP at some point $\left(\mu_{B},\kappa_{B}\right)$.
Near $\mu_{B}$ we express generically ${\cal R}\left(\kappa;\mu\right)\sim a\left(\mu\right)\kappa+b\left(\mu\right)\kappa^{2}+c\left(\mu\right)\kappa^{3}$
because the need for a strong-coupling solution $\kappa_{\infty}^{0}=0$
prevents a constant term and requires at least a cubic form to achieve
BP. With generically analytic coefficients at $\mu_{B}$, we expect  to leading order(s)
$a\left(\mu\right)\sim a_{0}+a_{1}\left(\mu-\mu_{B}\right)$, $b\left(\mu\right)\sim b_{0}$,
$c\left(\mu\right)\sim c_{0}$ for $\mu\to\mu_{B}$. Locating BP at
$\left(\mu_{B},\kappa_{B}\right)$ fixes $a_{0}=1+c_{0}\kappa_{B}^{2}$
and $b_{0}=-2c_{0}\kappa_{B}$. To orient BP correctly requires $a_{1}/c_{0}<0$,
and we set $a_{1}=-c_{0}A^{2}$ with $A>0$. Finally, stability of
the strong-coupling FP at $\kappa_{\infty}^{0}$ demands $c_{0}<0$,
and we may set $c_{0}=-1$. This yields 
\begin{equation}
\kappa_{n+1}-\kappa_{n}\sim\frac{\Delta\kappa}{\Delta n}\sim\left[-\kappa_{B}^{2}+A^{2}\left(\mu-\mu_{B}\right)\right]\kappa_{n}+2\kappa_{B}\kappa_{n}^{2}-\kappa_{n}^{3}\label{eq:Landau}
\end{equation}
as a minimal model. After extracting $\kappa_{\infty}^{0}\equiv0$,
the remaining FP equation indeed produces  by design a BP at $\left(\mu_{B},\kappa_{B}\right)$
with FP-branches 
\begin{equation}
\kappa_{\infty}^{\pm}=\kappa_{B}\pm A\sqrt{\mu-\mu_{B}}\label{eq:branchcut}
\end{equation}
for $\mu>\mu_{B}$. Local expansion near each FP as in Eq.~(\ref{eq:linearRG})
provides the eigenvalues $\lambda\left(\mu\right)=\partial_{\kappa}{\cal R}\left(\kappa_{\infty};\mu\right)$
depicted in Fig.~\ref{fig:lambda}(a), 
\begin{eqnarray}
\lambda^{0} & = & 1-\kappa_{B}^{2}+A^{2}\left(\mu-\mu_{B}\right),\nonumber \\
\lambda^{\pm} & = & 1\mp2A\kappa_{B}\sqrt{\mu-\mu_{B}}-2A^{2}\left(\mu-\mu_{B}\right),\label{eq:lambda}
\end{eqnarray}

\section{Discussion of the RG-Regimes}
\begin{figure}
\includegraphics[bb=45bp 30bp 350bp 350bp,clip,width=0.53\columnwidth]{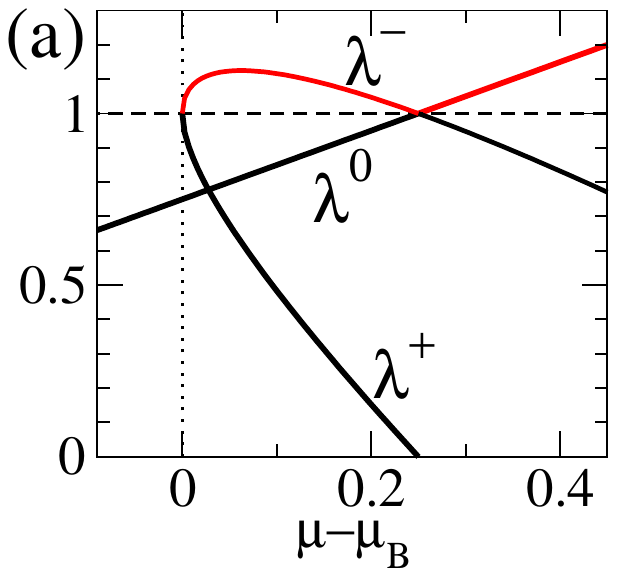}\hfill{}\includegraphics[clip,width=0.45\columnwidth]{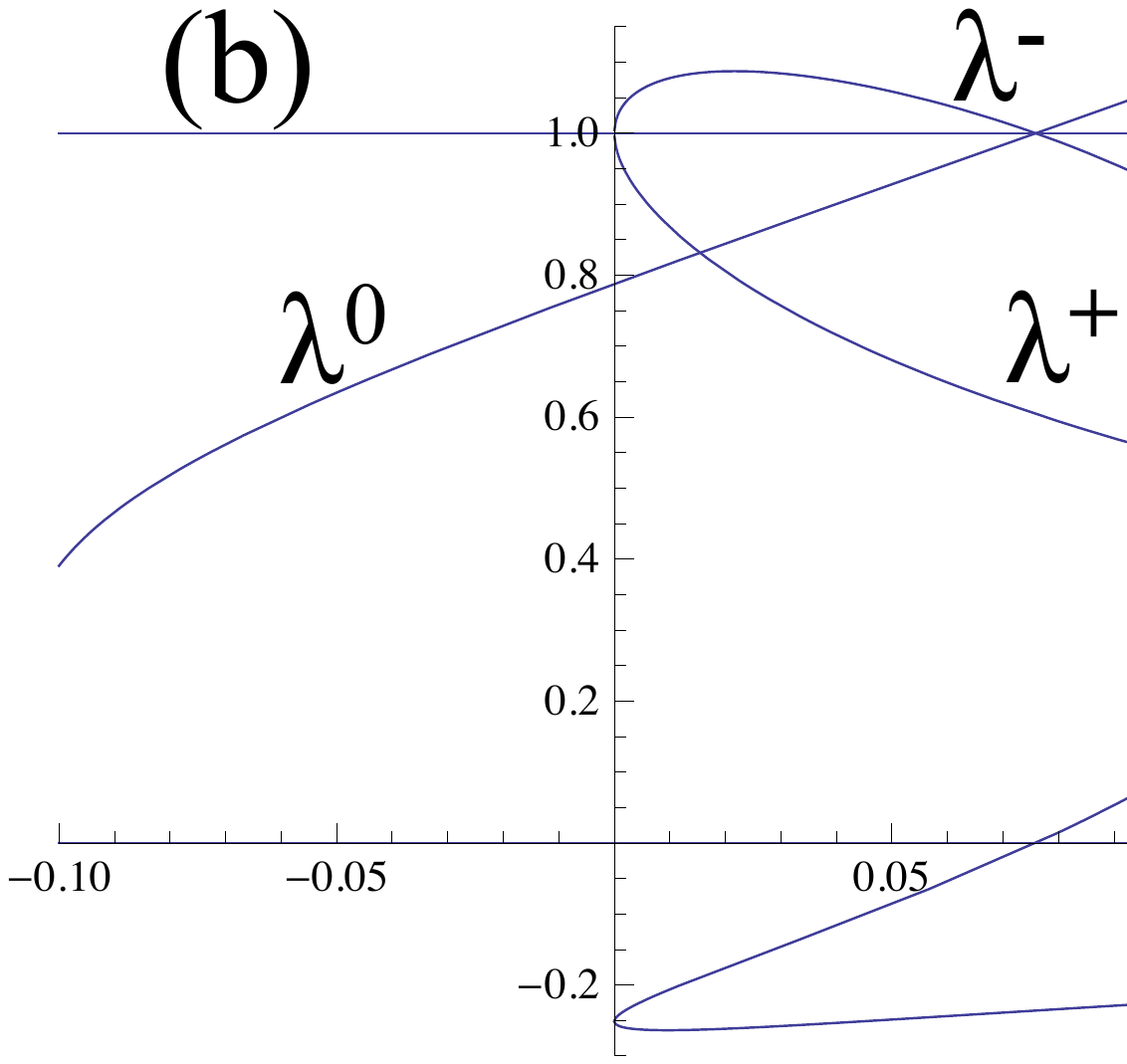}

\protect\caption{\label{fig:lambda}Plot of the eigenvalues (a) in Eqs.~(\ref{eq:lambda})
for the RG-flow in Eq.~(\ref{eq:Landau}), here $A=1$, $\kappa_{B}=\frac{1}{2}$,
and (b) for the RG-flow in Ref.~\cite{Boettcher10c} for HN5 (at
$y=0.1$). At BP ($\mu=\mu_{B}$), two conjugate eigenvalues emerge
simultaneously with marginal stability, $\lambda^{\pm}\left(\mu_{B}\right)=1$,
such that $\lambda^{-}$ remains unstable ($>1$, marked red) until
the lower FP-branch, $\kappa_{\infty}^{-}$, drops below $\kappa_{\infty}^{0}$,
leading to an intersection of $\lambda^{0}$ and $\lambda^{-}$ at
some $\mu>\mu_{B}$. For physical $\kappa_{B}>0$, the critical point
occurs at BP, $\mu_{c}=\mu_{B}$, while for unphysical $\kappa_{B}<0$,
setting $\kappa_{B}\to-\kappa_{B}$ merely swaps $\lambda^{\pm}\to\lambda^{\mp}$
(see Eq.~(\ref{eq:lambda})), and now $\lambda^{0}\left(\mu_{c}\right)=\lambda^{+}\left(\mu_{c}\right)=1$
intersect at $\mu_{c}>\mu_{B}$ with marginal stability. }
\end{figure}

In Fig.~\ref{fig:HN5yPlots}, panels (b) and (c) correspond to cases
where BP at ($\mu_{B},\kappa_{B}$) is in the physical domain ($0\leq\mu,\kappa\leq1$);
panel (d) represents $\kappa_{B}\leq0$. Within the domain of physical
$\kappa_{B}>0$, the lower FP-branch $\kappa_{\infty}^{-}\left(\mu\right)$
is unstable near $\mu_{B}$ ($\lambda^{-}>1$, see Fig.~\ref{fig:lambda})
while $\kappa_{\infty}^{+}\left(\mu\right)$ remains stable. Stable
and unstable branches merge at BP, where particularly interesting
phenomena arise. The decisive difference between panels (b) and (c)
is the location of BP relative to the initial $\kappa_{0}\left(\mu\right)$.

For the case of panel (b) (e.g., when long-range, hierarchical couplings
are weakest~\cite{Boettcher10c}), $\mu_{B}$ is small and/or $\kappa_{B}$
is closer to unity (or even above). Then, $\kappa_{0}\left(\mu\right)$
merely intersects the unstable branch $\kappa_{\infty}^{-}\left(\mu\right)$
at some critical point $\mu_{c}>\mu_{B}$ defined by $\kappa_{\infty}^{-}\left(\mu_{c}\right)=\kappa_{0}\left(\mu_{c}\right)$.
The RG-flow (vertical blue arrows in Fig.~\ref{fig:HN5yPlots}) for
$0\leq\mu<\mu_{c}$ advances toward strong coupling, $\kappa_{\infty}^{0}$,
while for $\mu_{c}<\mu\leq1$ it flows toward $\kappa_{\infty}^{+}\left(\mu\right)$
\footnote{Note that far away from $\mu_{B}$, $\kappa_{\infty}^{+}\left(\mu\right)\to1$
only for some $\mu>\mu_{c}$, reflecting the physical phenomenon of
``patchiness''~\cite{Boettcher09c,Boettcher11d}: hierarchical,
long-range couplings enforce some semblance of order between otherwise
uncorrelated (sub-extensive) patches of locally connected degrees
of freedom even in the disordered regime; full disorder is often only
reached at infinite temperature, dilution, etc (i.e., $\mu\to1$).%
}. Near $\mu_{c}$, the critical dynamics of the system is now determined
by the local properties of the unstable FP $\kappa_{\infty}^{-}\left(\mu_{c}\right)$
that has been selected by the specific system via its  bare coupling
$\kappa_{0}\left(\mu\right)$. As for a conventional system in Eq.~(\ref{eq:linearRG}),
local analysis~\cite{Pathria} of Eq.~(\ref{eq:Landau}) near $\kappa_{\infty}^{-}\left(\mu_{c}\right)$
yields the diverging correlation length, 
\begin{equation}
\xi\sim\left|\mu-\mu_{c}\right|^{-\nu\left(\mu_{c}\right)},\qquad\mu\to\mu_{c},\label{eq:xi_nu}
\end{equation}
but with a non-universal  thermal exponent $y_{t}=\log_{2}\lambda^{-}\left(\mu_{c}\right)=1/\nu\left(\mu_{c}\right)$.
For $\mu_{c}\searrow\mu_{B}$, $\lambda^{-}$ becomes marginal and
the exponent diverges as $\nu\left(\mu_{c}\right)\sim1/\sqrt{\mu_{c}-\mu_{B}}$.
Yet, for $\mu_{c}>\mu_{B}$, the RG-flow \emph{never} passes sufficiently
near BP.

For the case of panel (c) where $\kappa_{0}\left(\mu\right)$ passes
above BP (e.g., for somewhat stronger long-range couplings~\cite{Boettcher10c}),
the RG-flow \emph{must} pass BP which now dominates criticality, i.e.,
$\mu_{c}=\mu_{B}$, with an infinite-order divergence characterizing
this regime. Well below (above) $\mu_{B}$, the RG-flow evolves unperturbed
to $\kappa_{\infty}^{0}$ (to $\kappa_{\infty}^{+}$), the closest
stable FP. However, just below $\mu_{B}$ the RG-flow gets ever more
impeded near BP before it can reach $\kappa_{\infty}^{0}$. Asymptotically
for $\mu\nearrow\mu_{B}$ near $\kappa_{n}\sim\kappa_{B}+\epsilon_{n}$
with small $\epsilon_{n}$ at large but intermediate $n$, Eq.~(\ref{eq:Landau})
provides 
\begin{equation}
\epsilon_{n+1}-\epsilon_{n}=\frac{\Delta\epsilon_{n}}{\Delta n}\sim-\kappa_{B}A^{2}\left(\mu_{B}-\mu\right)-\kappa_{B}\epsilon_{n}^{2}.\label{eq:KT}
\end{equation}
This relation exhibits a boundary layer, i.e., in the limit $\mu\nearrow\mu_{B}$
the solution drastically changes behavior. With the methods of Ref.~\cite{BO},
we rescale $\epsilon_{n}\to\gamma\epsilon_{n}$ and $n\to\delta n$
to obtain a balance for $\delta\sim1/\gamma\sim1/\sqrt{\mu_{B}-\mu}$.
Accordingly, the characteristic width of the boundary layer scales
with $n^{*}\sim1/\sqrt{\mu_{B}-\mu}$, which leads to the divergence
in the correlation length characteristic of BKT, 
\begin{equation}
\xi\left(\mu\right)\sim2^{n^{*}}\sim e^{\frac{const}{\sqrt{\mu_{B}-\mu}}},\qquad\mu\nearrow\mu_{B}=\mu_{c}.\label{eq:xi_BKT}
\end{equation}
Clearly, the physical origin of this singularity is not related to
an actual BKT transition, with its formation of delicate topological
structures, that requires a rare confluence of dimensionality and
internal degrees of freedom for lattice models~\cite{Plischke94}.
In fact, instead of being rare, it appears as one of three generic
types of transition often found in hierarchical networks~\cite{Dorogovtsev08}.

The most unconventional behavior is depicted in panel (d) of Fig.~\ref{fig:HN5yPlots},
when $\kappa_{B}<0$ and BP has dropped below the physical regime,
corresponding to the situation when long-range couplings dominate~\cite{Boettcher10c}).
No unstable FP can be reached for any physical choice of $\kappa_{0}\left(\mu\right)$.
The RG-flow always advances to the closest stable FP, either at strong
coupling, $\kappa_{\infty}^{0}$ for $0\leq\mu<\mu_{c}$, or at patchy
order, $\kappa_{\infty}^{+}\left(\mu\right)$ for $\mu_{c}<\mu\leq1$.
Both lines of FP cross at $\mu_{c}(>\mu_{B})$, defined by the intersection
$\kappa_{\infty}^{+}\left(\mu_{c}\right)=\kappa_{\infty}^{0}\equiv0$.
This condition implies that both their eigenvalues are simultaneously
equal \emph{and} marginal, $\lambda^{0}\left(\mu_{c}\right)=\lambda^{+}\left(\mu_{c}\right)=1$,
as $\kappa_{\infty}^{0}$ must invert its stability at the intersection,
making marginal stability inherent to \emph{any} such system. In our
model, $\kappa_{\infty}^{+}\left(\mu_{c}\right)=0$ in Eq.~(\ref{eq:branchcut})
provides $-\kappa_{B}=A\sqrt{\mu_{c}-\mu_{B}}$, hence, Eqs.~(\ref{eq:lambda})
give $\lambda^{0,+}\sim1\pm A^{2}\left(\mu-\mu_{c}\right)$ for $\mu\to\mu_{c}$,
see Fig.~\ref{fig:lambda}(a). Near $\kappa_{\infty}^{0,+}\left(\mu_{c}\right)=0$,
the local analysis on Eq.~(\ref{eq:Landau}) according to Eq.~(\ref{eq:linearRG})
gives $\epsilon_{n+1}\sim\lambda^{0,+}\epsilon_{n}$ or $\epsilon_{n}\sim\epsilon_{0}\exp\left(-nA^{2}\left|\mu-\mu_{c}\right|\right)$
with a cross-over at $n^{*}\sim1/\left|\mu-\mu_{c}\right|$ that is
generic when $\lambda^{0}$ and $\lambda^{+}$ intersect linearly.
Thus, the divergence is 
\begin{equation}
\xi\left(\mu\right)\sim2^{n^{*}}=e^{\frac{const}{\left|\mu-\mu_{c}\right|}},\qquad\mu\to\mu_{c}\label{eq:def_xi-1}
\end{equation}
for the correlation length. Again, the RG flow does not pass BP, since
it is located \emph{below} the physical domain.

\section{Behavior of the Order Parameter} 
We can extend the discussion to include the effect of further control
parameters, such as an external field $\eta_{0}=\eta=e^{-\beta h}$.
Its generic RG-flow can be expressed asymptotically as
\begin{equation}
\eta_{n+1}\sim\eta_{n}^{\lambda_{h}},\qquad\lambda_{h}\sim2-C\kappa_{\infty}^{+}\left(\mu_{c}\right)\label{eq:etaRG}
\end{equation}
with some $\mu_{c}$-dependent constant $C>0$, near the critical
point $\mu_{c}$ and for sufficiently small $\kappa_{\infty}^{+}\left(\mu_{c}\right)$.
This satisfies the physical requirements on its FP, $\eta_{\infty}=0,1$;
only for $h=0$ the RG-flow remains at the unstable FP, i.e., $\eta_{n}\equiv1$
f.~a.~$n$, and for any $h>0$ the stable strong-coupling FP at
$\eta_{\infty}=0$ is reached eventually. The eigenvalue near the
unstable FP satisfies $\lambda_{h}\leq2$ such that the  magnetic exponent becomes $y_{h}=\log_{2}\lambda_{h}\leq1$,
as shown in Ref.~\cite{PhysRevLett.108.255703}. There, a scaling
theory is developed concerning the BKT regime based on the exponentially
divergent correlation length in Eq.~(\ref{eq:xi_BKT}), leading to
an order-parameter (magnetization, fraction of sites on percolating
cluster, etc.) 
\begin{equation}
m\sim\xi^{-1}\sim\exp\left\{ -\frac{const.\left(1-y_{h}\right)}{\left(\mu_{c}-\mu\right)^{-x_{t}}}\right\} ,\quad\mu\nearrow\mu_{c},\label{eq:mBKT}
\end{equation}
when $y_{t}\to0$. Of course, for $y_{t}>0$ it is~\cite{Pathria}
\begin{equation}
m\sim\left(\mu_{c}-\mu\right)^{\beta},\qquad\beta=\frac{1-y_{h}}{y_{t}}\label{eq:beta}
\end{equation}
for small-world systems where $N$ takes the role of $L^{d}$~\cite{PhysRevLett.108.255703}.

\begin{figure}
\includegraphics[clip,width=0.3\columnwidth]{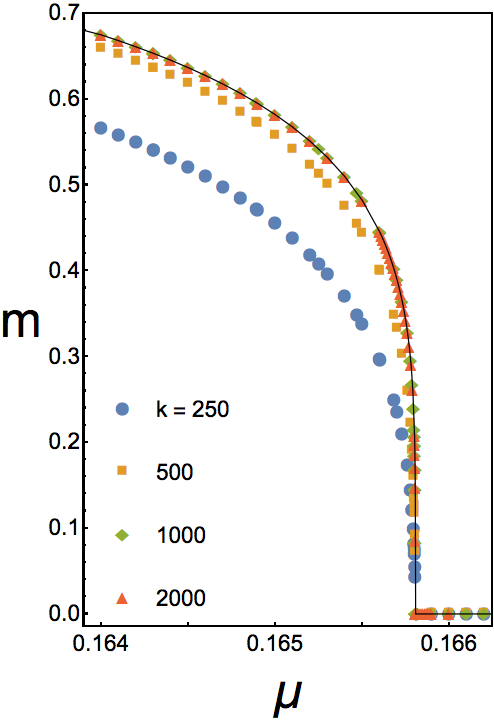}\includegraphics[bb=0bp 0bp 400bp 550bp,clip,width=0.33\columnwidth]{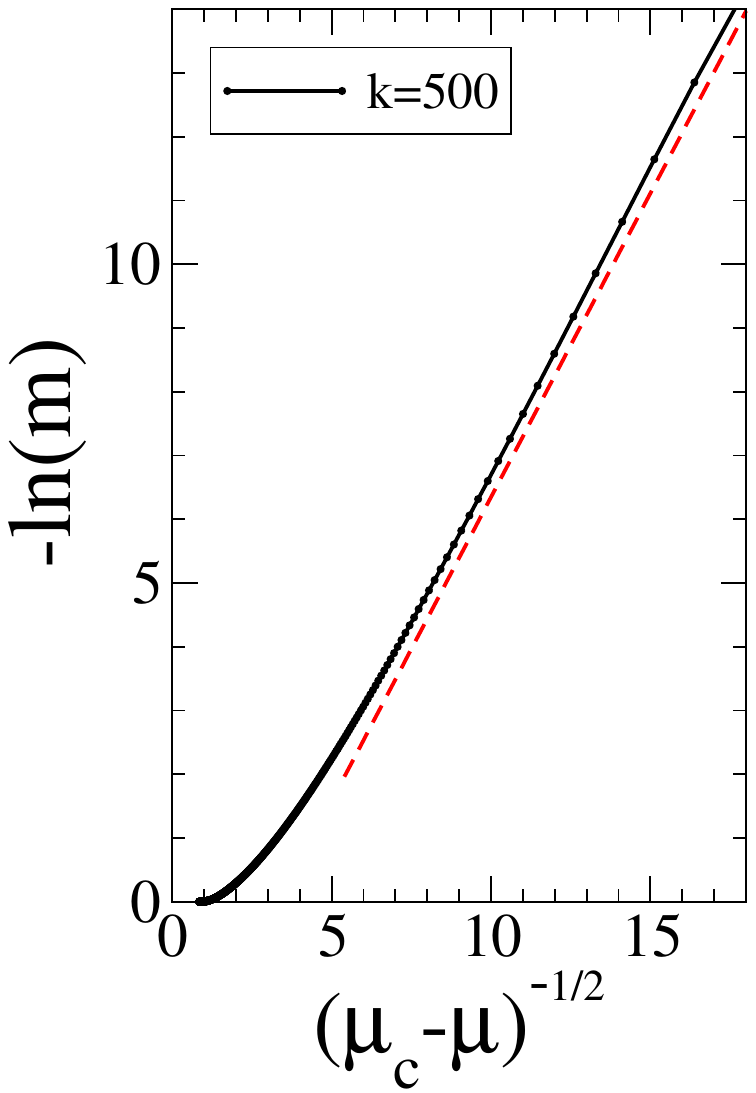}\includegraphics[clip,width=0.33\columnwidth]{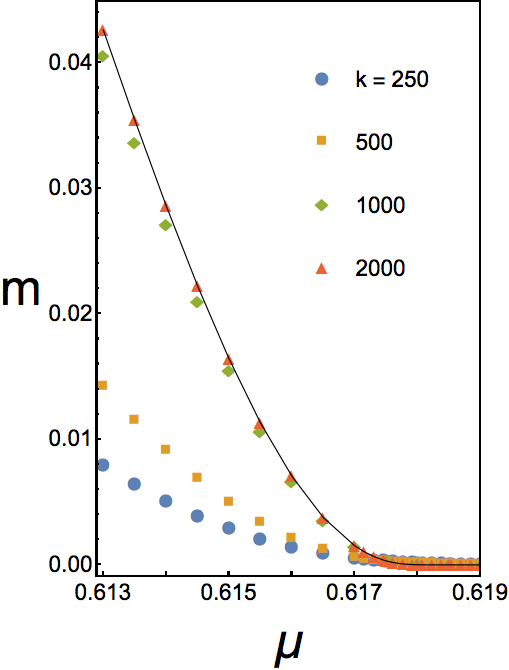}

\protect\caption{\label{fig:mBKT} Plot of the magnetization in the Ising model on
the hierarchical network HN5 from Ref.~\cite{Boettcher10c}  with balance $y=K/L$ between nearest-neighbor and long-range couplings, $K$ and $L$, (a) for
$y=0.1$, (b) for $y=0.4$, and (c) for $y=1$, which in sequence
correspond to regimes (b) to (d) in Figs.~\ref{fig:HN5yPlots}. The
transition is (a) \emph{2nd}-order continuous with $\beta=0.205\ldots$,
(b) BKT-like, and (c) again continuous with $\beta=\left(3+\sqrt{5}\right)/4=1.30\ldots$.
Strong finite-size effects ($N\sim2^{k}<\infty$) remain throughout.}
\end{figure}

Our theory in Eq.~(\ref{eq:Landau}) not only explains the robustness
of the value of $x_{t}=\frac{1}{2}$ conjectured in Ref.~\cite{PhysRevLett.108.255703},
but also broadens the scope to a total of three generic regimes in
the divergence of $\xi$, as we have explained. With the addition
of Eq.~(\ref{eq:etaRG}), we can account for the behavior of the
order-parameter $m$. For the first two regimes where $\kappa_{B}>0$,
it is $\lambda_{h}<2$ in Eq.~(\ref{eq:etaRG}) so that $1-y_{h}>0$.
In the regime with the weakest distortion of the FP, see Fig.~\ref{fig:HN5yPlots}(b),
we have shown in Eq.~(\ref{eq:xi_nu}) that $y_{t}=1/\nu\left(\mu_{c}\right)>0$,
i.e., it is $0<\beta\left(\mu_{c}\right)<\infty$, similar to an ordinary
2\emph{nd}-order transition, except for its non-universal $\mu_{c}$-dependence.
In the BKT-regime, see Fig.~\ref{fig:HN5yPlots}(c), it is $\mu_{c}=\mu_{B}$
and $\lambda^{+}\left(\mu_{c}\right)=1$ in Eq.~(\ref{eq:lambda})
such that $y_{t}=0$ and $\beta\to\infty$, which leads to Eq.~(\ref{eq:mBKT})
described in Ref.~\cite{PhysRevLett.108.255703}. Finally, when $\kappa_{B}<0$,
see Fig.~\ref{fig:HN5yPlots}(d), it is $\mu_{c}>\mu_{B}$ and $\kappa_{\infty}^{+}\left(\mu_{c}\right)=0$.
Then, $\lambda^{+}\left(\mu_{c}\right)\to1$ such that $y_{t}=O\left(\mu_{c}-\mu\right)$
leads to the divergent correlation length in Eq.~(\ref{eq:def_xi-1}),
however, it is $\lambda_{h}\to2$ in Eq.~(\ref{eq:etaRG}) such that
$y_{t}=1-o\left(\mu_{c}-\mu\right)$. In the most generic (analytic)
case, we would expect that both, $y_{t}$ and $y_{h}$, have linear
corrections so that $\beta$ in Eq. \eqref{eq:beta} remains positive
for $\mu\to\mu_{c}$ and the transition is continuous. This is indeed
the observed phenomenology, for instance, for the one-parameter family
of Ising models~\cite{BrBo14}, first studied in Ref.~\cite{Boettcher10c},
that interpolates between $\kappa_{B}>0$ and $\kappa_{B}<0$. Surprisingly,
percolation models on these hierarchical networks appear to provide
quite common exceptions to this behavior~\cite{Boettcher11d,Nogawa13,Singh14,Singh14b},
with $\beta=0$, resulting in a remarkable \emph{discontinuous} (``explosive'')
percolation transition~\cite{Cho14}. In Ref.~\cite{Singh14b}, it
was argued that such non-generic behavior, in form of merely a $2nd$-order
correction in $y_{h}$ throughout these models, originates with the
interplay of tree-like (hyperbolic) features superimposed on a geometric
($1d$-lattice) structure common to those networks. 

As a demonstration for our theory, in Ref.~\cite{BrBo14} we revisit
the Ising model on the Hanoi network HN5 previously considered for
$h=0$ in Ref.~\cite{Boettcher10c}. There, a one-parameter family
of Ising models was conceived via the ratio between short-range and
small-world coupling strengths that interpolates between all three
regimes; Fig.~11 in Ref.~\cite{Boettcher10c} corresponds to Fig.~\ref{fig:HN5yPlots}(b-d)
here. That system is far more complex than our model here in that
there are two couplings and three fields (when $h>0$, for site-,
bond-, and three-point magnetizations) to be renormalized. Yet, the
same three regimes in the divergence of the correlation length $\xi$
and the magnetization $m$ ensue, as our theory predicts. Here, we
only plot the magnetization of the Ising model on a hierarchical network,
which follows 2nd-order behavior, Eq.~(\ref{eq:beta}), in the first
regime in Fig.~\ref{fig:mBKT}(a), it has an infinite-order transition,
Eq.~(\ref{eq:mBKT}), in the BKT regime in Fig.~\ref{fig:mBKT}(b),
and it becomes again continuous in the regime of intersecting stable
FP in Fig.~\ref{fig:mBKT}(c).

\section{Conclusions:}
We have introduced a simple RG-model to categorize the regimes of synthetic critical behaviors in hierarchical networks.  The robustness of these regimes derives from
the fact that  branch points in control-parameter dependent RG-flows are most generically a square-root singularity. Our theory specifically addresses the question~\cite{PhysRevLett.108.255703} about
the universality of the BKT result in Eq.~(\ref{eq:xi_BKT}). The full exponential singularity in
Eq.~(\ref{eq:def_xi-1}) is even more robust, as it does not depend
on the nature of  the branch point singularity but merely in the fact that two intersecting lines
of fixed points must switch stability. This \emph{implies} marginally stable eigenvalues
at the point of intersection. Those eigenvalues invariably scale linearly
with the control parameter there. For the future, it would be interesting to explore our model prediction directly for hierarchical networks drawn from some ensemble, instead of exactly renormalizable instance. 

This work was supported by DMR-grant \#0812204 from the NSF. SB would
like to thank T. Nogawa, T. Hasegawa, N.~Berker and P.~Phillips
for helpful discussions.  

\bibliographystyle{eplbib}
\bibliography{/Users/stb/Boettcher}

\end{document}